# Spatio-temporal variation of temperature for the recent 40 years in Lhasa

Yang Lei[1#], Hongbin Wang[1#], Tongsuo Lu[1,2], Wenxue Fu[1], Jing Hu[1], Dong Wei[3]

[1] College of Science, Tibet University, Lhasa, 850000 China

[2] Shanghai Institute of Applied Physics, Chinese Academy of Sciences, Jiading, Shanghai 201800, China

[3] Lhasa Meteorological Bureau, Lhasa, 85000, China

**Correspondence**

Tongsuo Lu, College of Science, Tibet University, Lhasa, 850000, China or Shanghai Institute of Applied Physics, Chinese Academy of Sciences, Jiading, Shanghai 201800, China.

E-mail: lutongsuo@163.com

[#]Equal contribution, co-first author.

**Funding information**

National Natural Science Foundation of China, Grant/Award Number: 11803024,11747128; Natural Science Foundation of Tibet, China, Grant/Award Number: XZ2019ZRG-163.

**Abstract:** It was all known that Lhasa went through a high temperature of 30.8°C in late June 2019, which hit record highs. To better understand the reasons, based on observations recorded at automatic weather stations in Lhasa, we studied the characteristics of temperature variation at multiple time scales using the linear trend method, Mann-Kendall mutation test, morlet wavelet analysis, R/S analysis and so on. The results showed that: (a) The annual mean temperature (AMT) is rising at a rate of 0.5°C/10yr, and the average temperature for different seasons also increased significantly, especially in winter. (b) Although there was an intersection in 1995, we found that AMT, did not pass the reliability test of significance level α=0.05, this means there are no abrupt changes for AMT, the values are 7.97°C and 9.15°C respectively before and after the intersection point. (c) AMT has a periodic oscillation for 18~25yr and 25~32yr based on a mass of data and the wavelet variance diagrams in Lhasa. AMT has a main cycle of 28yr, cyclic Patterns of temperature changes in spring, summer and autumn is similar to AMT, but it is relatively complex in winter. (d) The Hurst index of AMT and different seasons demonstrates that the temperature are likely to continue to rise in the future in Lhasa.

# 1. INTRODUCTION

As per Intergovernmental Panel on Climate Change (IPCC), the anthropogenic greenhouse gas emissions have increased since the pre-industrial era (IPCC, 2014), which has led to the warming of the Earth's atmosphere (Karl and Trenberth, 2003). The rising temperature led to alterations in climatic patterns around the globe (Trenberth, 2011; Ashutosh S., et al., 2020). Global warming has become the focus of the world's attention in recent years. Based on current knowledge of climate change, the need to study its regional and local processes more accurately is more than evident since this is precisely where most uncertain aspects are found.( Miró., et al., 2015) An in-depth analysis of the climate parameters, i.e., temperature, which essential for gaining an understanding of the recent-past and present climate of a region Almazroui et al., (2012). As one of the most important climatic factors. Temperature changes have greatly affected the normal production and life of human beings. The fifth assessment report of the intergovernmental panel on climate change (IPCC) clearly points out years Barros and Stocker (2012). The global mean surface temperature of sea and land has an obvious tendency to rise from 1880 to 2012, increased by 0.85°C (0.65~1.06°C). The rise rate (0.12°C/10 yr) between 1951 to 2012 was twice as fast as it has been since 1880 Qin and Thomas (2014). Many scholars (e.g. Zhang et al., 2006; Wang et al., 2014; LI et al., 2012; Zhang et al., 2014; Kong 2010) have found that the rise rate is 0.25°C/10a in China, AMT has increased by1.3°C. However, due to the space-time characteristics of the temperature change is not the same in change is not the same in different regions. Monitoring temperature changes is conducive to better understanding and evaluating the causes and trends of global warming in some areas Sun et al., (2007). It was all known that Lhasa went through a high temperature of 30.8°C in late June 2019, which hit record highs. Lhasa is located in the central part of the Tibetan plateau, with an average altitude of 3,658 meters. And it is the political, economic, cultural, scientific and educational center of Tibet, as many as one million people live there, Tibet plateau is known as the "third pole of the earth" with a specific natural environment, special life condition, harsh climate and extremely fragile ecological environment, it is also essentially the source point of all of Asia's major rivers. The results of our research will help local residents, workers sent to support Tibet, builders, scientists and engineers better adapt to the living environment. Also to provide references and some constructive suggestions for decision-making departments to establish the warning mechanisms of high and low temperature, as well as make public health measures for the population and protect the local ecological environment under the big background of global warming.

## 2. DATA SOURCE

We selected some observation data at automatic weather stations in Lhasa (provided by Lhasa meteorological bureau) as the credible data source for this research，the time series is from January 1978 to December 2018. Daily mean temperature (DMT), daily maximum temperature (DMAX) and daily minimum temperature (DMIN) are recorded in the data file. According to the principle of meteorology, we will divide the whole year into four seasons: spring (March to May), summer (June to August), autumn (September to November) and winter (December to the next February) Lai (1996), we retrieve and look at all the data which are complete and reliable. However, the data ended in December 2018, the meteorological data for that winter were missing.

## 3. METHODS AND RESULTS

3.1. Linear trend method & sliding mean method

We use the linear trend method to study upward or downward trend of climatic factors for a long time. Trend coefficients represents what extent an association between variables and time. It eliminates the influence of mean squared error of temperature on the size of linear regression coefficient. Since temperature is likely to change at any time, a linear equation is selected to fit the variable series model when analyzing the trend of spatial-temporal temperature change. The correlation coefficient between time series x and original variable series y was used for the significant test (ɑ=0.05), as a result, only when the trend coefficient is greater or equal to the statistical significance criterion. Such a change will be considered as unpleasant change of climate that exceeds the normal rate of change.

We try to analyze and predict the tendency of temperature change with unitary linearity regression, Fig. 1 depicts the mean temperature (MT) is about 8.6 °C for the recent 40 years in Lhasa, the maximum AMT is 10.29°C appeared in 2009, the lowest AMT is 7.21 °C, appeared in 1978. A fluctuating and rising trend as shown in Fig. 1. A linear regression equation with unitary set up as below:

$$y=0.0591x + 7.3907$$

its rate of change is 0.591°C/10yr, the correlation coefficient between AMT and time series is r=0.86, which has passed the reliability test of significance level α=0.05, that means that AMT is rising significantly with a linear trend.

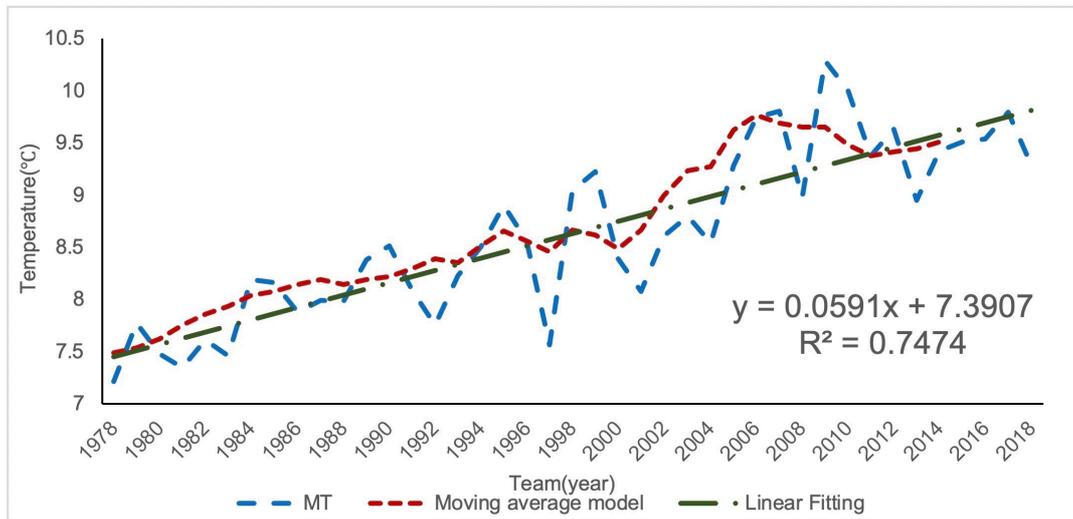

**Figure 1.** The variation of AMT in Lhasa in recent 40 years.

We can see that AMT covers three stages from the curve of moving average model (5yr): The cold period lasted from about 1978 to 1995, it is a period of warming and cooling cycles from 1996 to 2001, AMT fluctuates up and down in small range, and during 2002 to 2018 is a warmer period. Fig. 2 shows the mean temperature (MT) in different seasons for recent 40 years (spring, summer, autumn and winter) is 9.18°C, 16.03°C, 8.85°C and 0.43°C, respectively, the lowest and the highest seasonal mean temperatures (SMT) are respectively 7.2°C (in 1983) and 11.26°C (in 1999) in spring, the rate of change is 0.508°C/10yr. It is the similar to the others given in Table 1. In the end, we found that all of them have passed the test of significance level α=0.05, that means that SMT is rising significantly, as depicted in Fig. 2, an rising trend of fluctuation is presented.

**Table 1.** A summary of the mean temperature of the same season in different years (MT), the lowest SMT when it appears, the highest SMT when it appears, the rate of change and the correlation coefficient.

|  | MT (°C) | LT & when appear (°C/yr) | HT & when appear (°C/yr) | The rate (°C/10yr) | correlation coefficients (R) |
|---|---|---|---|---|---|
| Spring | 9.18 | 7.20, 1983 | 11.26, 1999 | 0.508 | 0.58 |
| Summer | 16.03 | 14.59, 1984 | 18.18, 2009 | 0.426 | 0.60 |
| Autumn | 8.85 | 6.85, 1986 | 10.79, 2017 | 0.645 | 0.80 |
| winter | 0.43 | -2.68, 1981 | 2.69, 2005 | 0.819 | 0.79 |

(Note: SMT is the abbreviation for "seasonal mean temperature", LT represents the lowest SMT, HT is. short for the highest SMT)

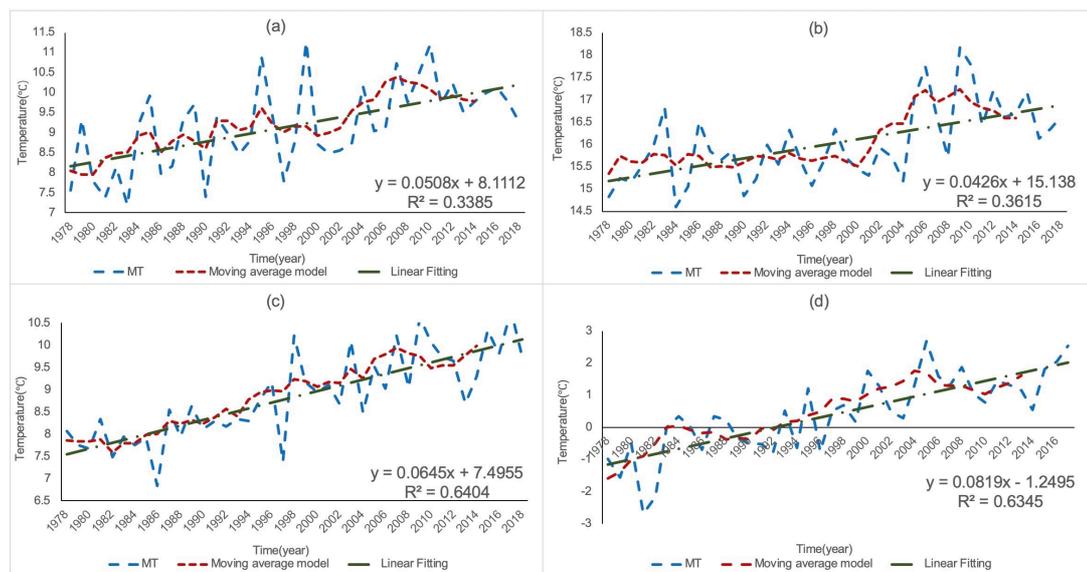

**Figure 2.** The temperature variation tendency of different seasons in Lhasa in recent 40 years.

[ (a) Spring   (b) Summer   (c) Autumn   (d) Winter ]

In the end, we found that all of them have passed the test of significance level α=0.05, that means that SMT is rising significantly, as depicted in Fig. 2, an rising trend of fluctuation is presented.

Sliding average, also called exponential weighted mean, is equivalent to low-pass filtering (LPF), which can be used to estimate the local mean value of a variable, so that the update of variables is related to the historical value over a period of time, and the smoothing value of time series is used to show the change of a certain element. We get the trend curves by moving average model with a period of about 5 yr, which are similar to the change trend of AMT.It can be divided into different periods in Fig 3.

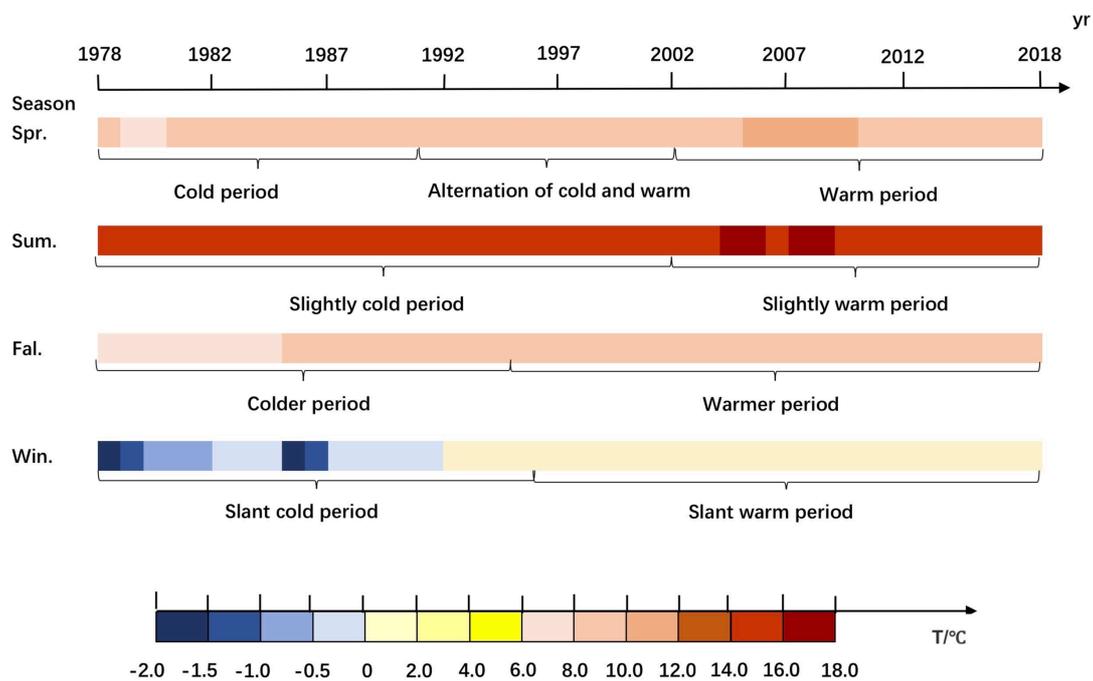

**Figure 3.** Different periods of cold and warm given by moving average model.

3.2. Mann-Kendall mutation test

Mann-Kendall is a simple and effective non- parametric test method (Stebor et al. 2016, Rovert et al. 1984). Its advantage is that the samples don't obey a certain distribution, and a minority of outliers do not affect the test results, which is easy to calculate.

We draw the curve chart of AMT and SMT by means of Mann-Kendall mutation test, as depicted in Fig. 4, there is a crossing point between the UF and UB curves in 1995, but it did not pass the reliability test of significance level α=0.05 (above the boundary line), this means there are no abrupt changes for AMT, the values are 7.97°C and 9.15°C respectively before and after the intersection point. Fig. 5 depicts that there is a crossing point between the UF and UB curves in 1994 for spring, but then three other crossing points were discovered behind it, to our surprise, these points are below the boundary line, that may also be interpreted as the proof of mutation for mean temperature in recent 40 years. The mean temperature shows an upward tendency after a slight oscillation around 1.96°C. In autumn and winter, each has a crossing point which is above the boundary line. At the same time, an abrupt change occurs in 2001 for summer which is below the boundary line.

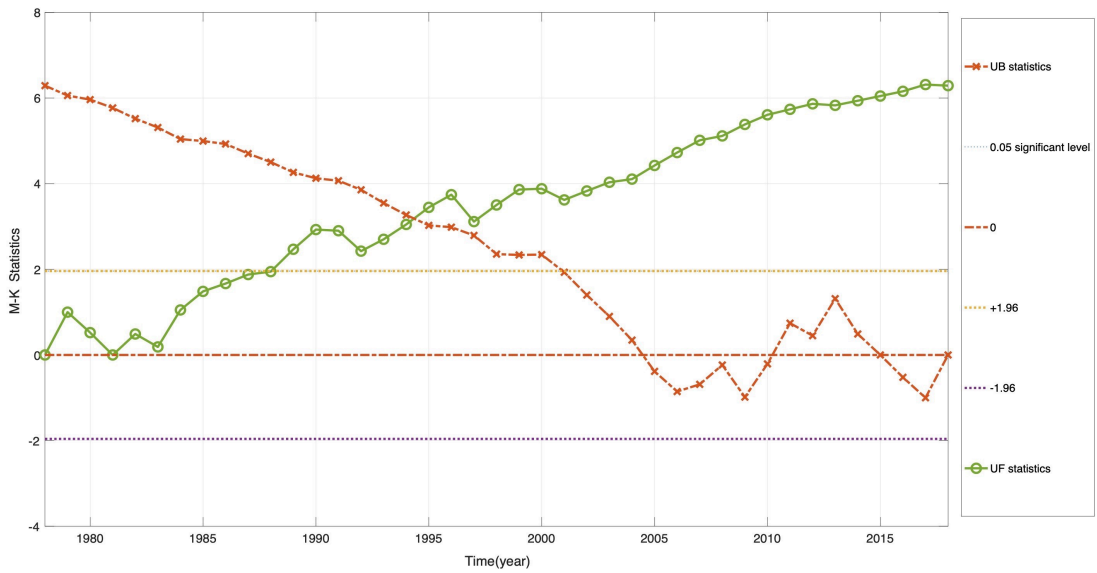

**Figure 4.** The curve chart of AMT by means of Mann-Kendall mutation test for recent 40 years.

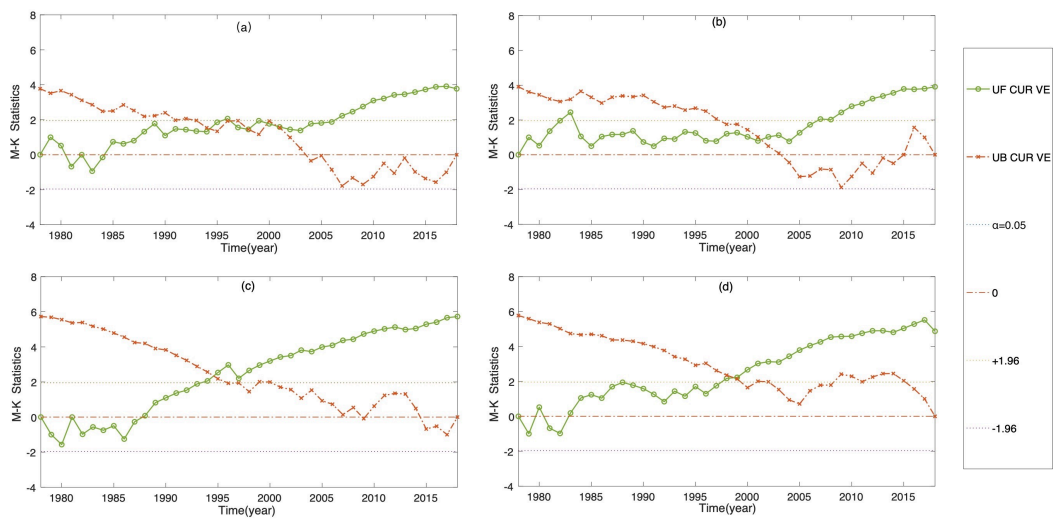

**Figure 5.** The curve chart of SMT by means of Mann-Kendall mutation test for recent 40 years.

[ (a) Spring   (b) Summer   (c) Autumn   (d) Winter ]

3.3. Morlet wavelet analysis

The wavelet analysis (WA) has local identification ability in time domain and frequency domain. The method of Morlet wavelet analysis is proposed to diagnose multi-level characteristics of temperature series variation, and thus the detailed information of periodic changes on time - scale is obtained.

Morlet wavelet was used to analyze the developing cycle of AMT in Lhasa, and the real part isogram of wavelet transform coefficients and wavelet variance chart were drawn in Fig. 6. The isogram of the real part shows a changeable climate between warmth and cold with different time scales. Different time scales are corresponding with the status of temperature change, many small scale changes are contained in large scale changes. The characteristics of oscillation period are relatively complex, as described in Fig. 6, different color shades and the size of the wavelet coefficients show signal strength, a positive number means warmer, the opposite means colder.

The map shows the specific changes over time in Table 2.

**Table 2.** The characteristics of temperature change with time - scale.

| period | <3yr | 3 ~ 10yr | 10 ~ 18yr | 18 ~ 25yr | 25 ~ 32yr |
|---|---|---|---|---|---|
| **The changing characteristics** | The change of AMT is simple, less mutations, extreme points are uniformly distributed. | signal frequent alternation but weak oscillation, the change of AMT is changeable but quite gentle. | signal frequent alternation but weak oscillation | extreme points are regular, alternative variation is clear and obvious, larger variable range | signal frequent alternation and intense oscillation, the change of AMT is obvious, a peak in 28a. |

Fig. 7a and 7b, shows that wavelet variance charts in spring and in summer are similar to the trend of AMT. There are two peaks (two primary periods) for autumn in 15yr and 28yr, respectively (Fig. 7c). As shown in Fig. 7d, SMT is complex and variable in winter. The temperature changes, mutations extreme points, fluctuating states and alternating signals are different in different stages. There are obviously four peaks (four primary periods) for winter in 6yr, 9yr, 15yr and 28yr, respectively. To a certain extent, SMT varies greatly for winter in

Lhasa.

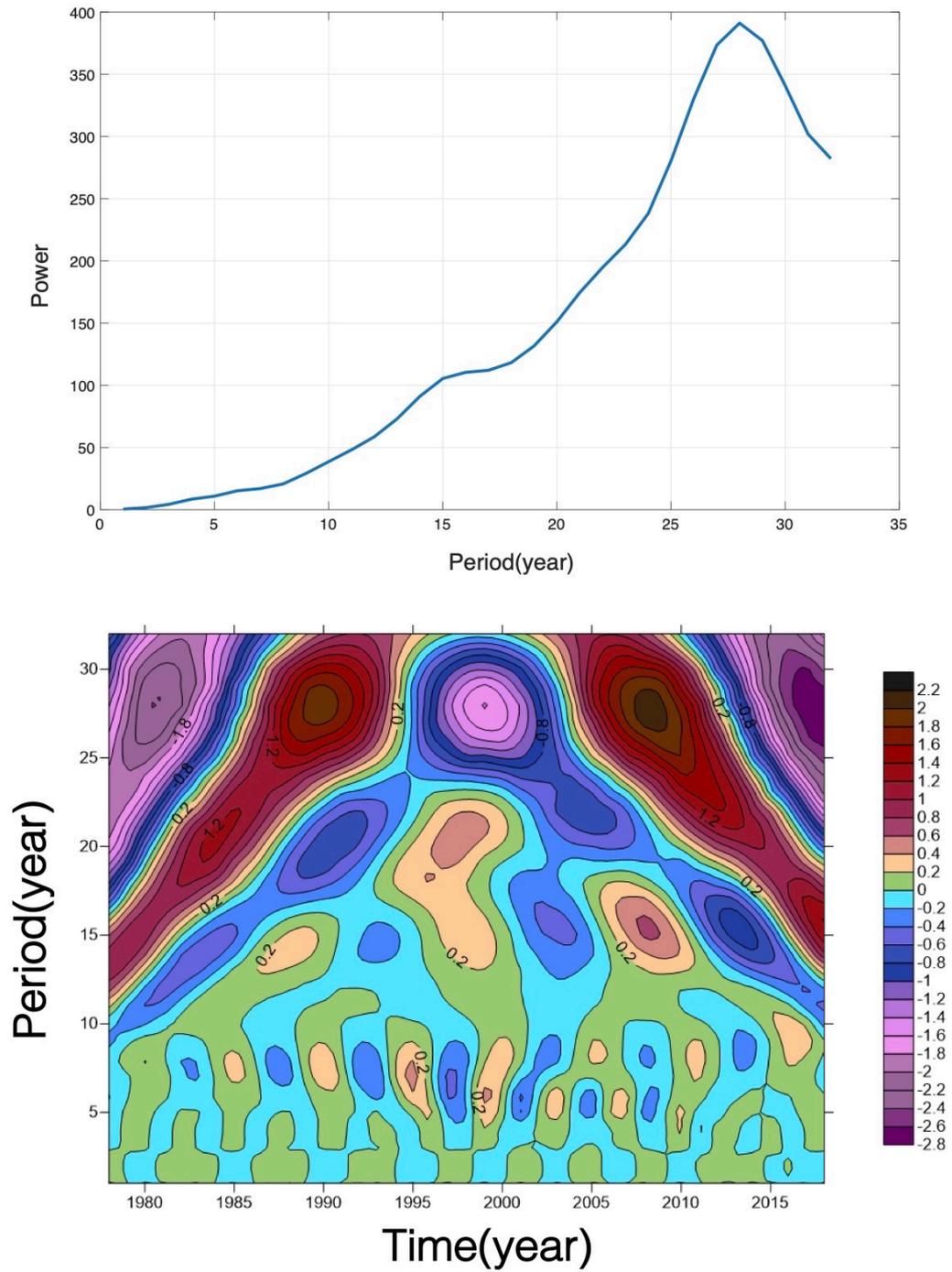

**Figure 6.**  Morlet wavelet analysis &wavelet variance chart of AMT.

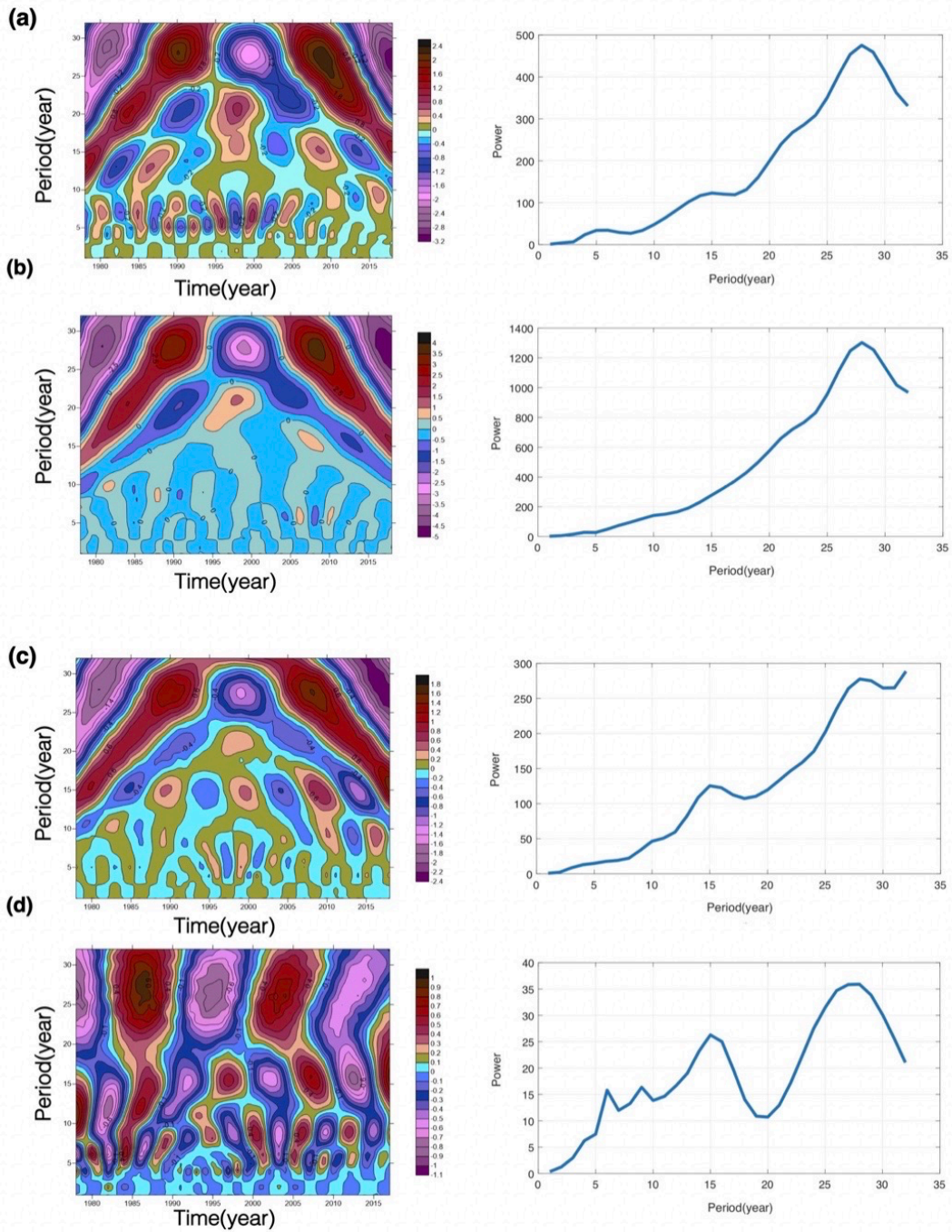

**Figure 7.** Morlet wavelet analysis &wavelet variance chart of SMT.

[ (a) Spring  (b) Summer  (c) Autumn  (d) Winter ]

3.4. R/S analysis

R/S analysis method was first proposed by British hydrologists named Hurts. Which is also known as the rescaled range analysis method. We can estimate the value of H by an algorithm of least square linear regression in a logarithmic graph, H is the index of Hurts Mandelbrot et al. (1969), and there are three situations as follows:

When 0≤H≤0.5, the time series is anti-persistent and has more significant. Mutation than random sequence. The general trend of temperature change is contrary to the past in the indicator of temperature.

When H=0.5, it means that the time series is a random walk sequence, and the observed results are completely independent and the temperature changes randomly in the indicator of temperature.

When 0.5<H≤1, the time series is persistent or trend-enhancing series which reflects the general trend of temperature change in the future is the same as in the past Zhao et al. (2001).

We got the index of Hurts (see Table 3.) with the methods of fitting the graph of the Hurts function of AMT and SMT. Based on this, we try to forecast the trend of temperature change in Lhasa, the results are shown in Fig. 8 and Fig. 9.

We can see that the values of the five sequences are greater than 0.5 (Although H=1.038 is slightly greater than 1 in autumn, there have no influence on the prediction of temperature change trend in winter). That shows, the change of AMT and SMT are positive correlated with time series. That means the general trend of temperature change in future is the same as in past, it is possible that the temperature is rising constantly and the weather continue to get warmer.

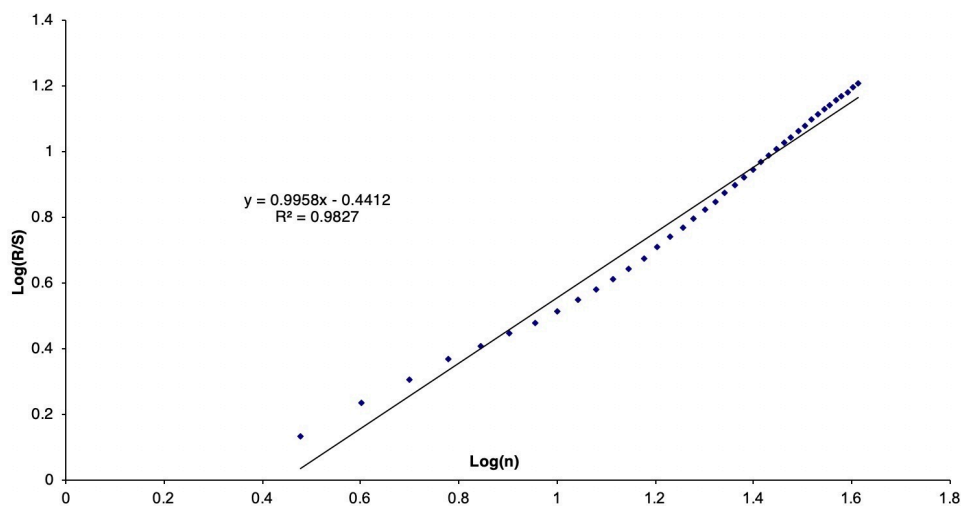

**Figure 8.** The index of Hurts of AMT based on R/S analysis method for recent 40 years.

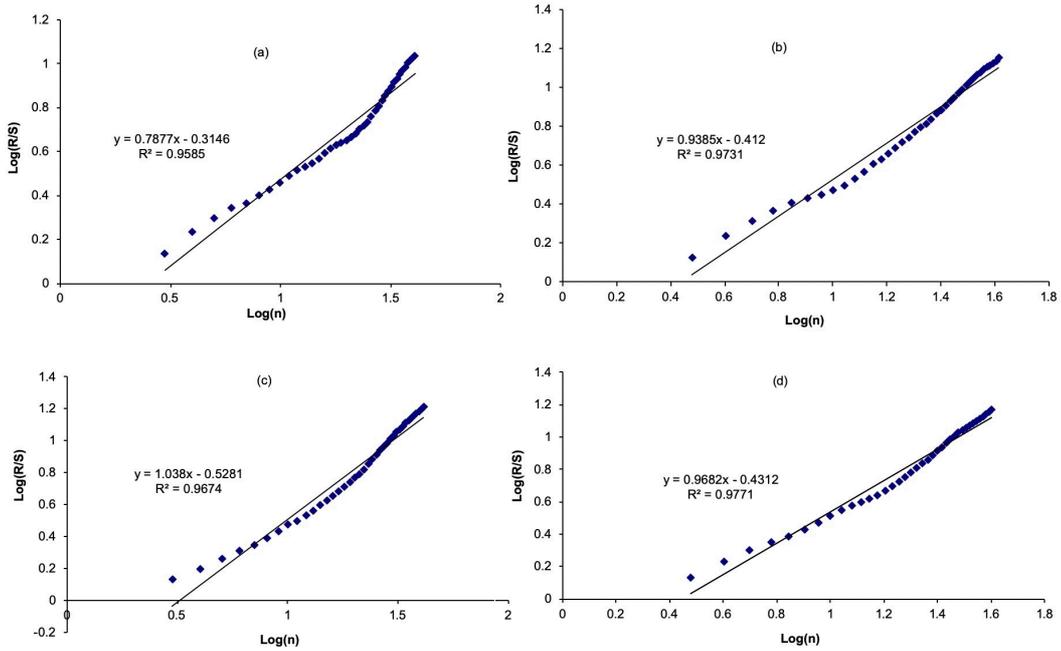

**Figure 9.** The index of Hurts of SMT based on R/S analysis method for recent 40 years.

[ (a) Spring   (b) Summer   (c) Autumn   (d) Winter ]

**Table 3.** The index of Hurts for AMT and SMT.

|         | spring | summer | autumn | winter | annual |
|---------|--------|--------|--------|--------|--------|
| Hurst(H)| 0.7877 | 0.9385 | 1.038  | 0.9682 | 0.9958 |

## 4. Conclusion and Discussion

The rising rate of AMT is 0.508°C/10yr, although SMT also rises, but slowly in summer and quickly in winter; autumn and winter are the main contributors of warming temperatures. We found that there was an intersection in 1995 by means of Mann-Kendall mutation test, nevertheless, AMT did not pass the reliability test of significance level α=0.05, this means there are no abrupt changes for AMT. SMT mutates more frequently in spring, after the abrupt change of temperature, it's getting warmer obviously. The principal period is about 28yr for AMT based on the method of Morlet wavelet analysis, there are four principal periods for winter and two principal periods for summer, respectively. Therefore the changes of temperature are relatively complex. The indices of Hurts of AMT and SMT are greater than 0.5. This shows, the change of AMT and SMT are positive correlated with time series.

That means the general trend of temperature change in future is the same as in past, it is possible that the temperature is rising constantly and the weather continue to get warmer.

Lhasa is on the Tibetan plateau, the ecological environment is very vulnerable, the climatic conditions are complex, the biodiversity is facing a serious threat, the self-adjustive ability of ecosystem is weak. With the rapid development of Tibetan economy and society, continuous increase in industrialization and urbanization. The indust-rial production, citizen life, vehicle exhaust pipes, coal plant smokestacks, burning trash piles are the main source of air pollution in Lhasa, AMT is rising in different extent on the back of global warming. In addition, there are many factors affect the change of temperature in Lhasa, for instance, the special topography, the large-scale pattern of atmospheric circulation, ground c[1] overing, and greenhouse gas emissions have also close relation with an abrupt change of temperature, which needs further research in the future.

## Acknowledgements

The authors would like to thank the National Natural Science Foundation of China (11803024,11747128) and Natural Science Foundation of Tibet，China (XZ2019ZRG-163) thanks cooperation units for their help.

---

[1] All authors declare no conflict of interest.